\documentclass[aps,prl,twocolumn,showpacs]{revtex4}
\usepackage{amssymb}
\usepackage{amsmath, amsthm}

\begin{document}


\title{A viability criterion for modified gravity with an extra 
force}


\author{Valerio Faraoni}
\email[]{vfaraoni@ubishops.ca}
\affiliation{Physics Department, Bishop's University\\
Sherbrooke, Qu\'ebec, Canada J1M~0C8
}


\begin{abstract} 
A recently proposed  theory of modified gravity with an explicit 
``anomalous'' coupling of the Ricci curvature to matter is 
discussed, and an inequality is derived which expresses a 
necessary and sufficient condition to avoid the notorius 
Dolgov-Kawasaki  instability.
\end{abstract}


\maketitle

\section{Introduction}
\setcounter{equation}{0}

Recently, modifications of gravity at cosmological scales have 
received much attention 
\cite{Capozzielloetal,CDTT, modifiedgravity}  in order to 
explain the cosmic acceleration discovered in 1998 using type 
Ia supernovae  \cite{SN}. The alternative is to resort to a 
mysterious form of 
dark energy  with exotic properties \cite{Linderresletter}: a 
negative 
pressure $P$ and an energy density $\rho$ satisfying $P\approx 
-\rho$ and perhaps even $P<-\rho$ (phantom energy 
\cite{phantom}). The latter easily leads to a Big Rip 
singularity at a finite future \cite{BigRip}. 
Modified gravity  allows one to avoid such exotica. The 
Einstein-Hilbert  action \footnote{Here ${\cal L}_m$ is the 
matter Lagrangian  density, $R$ is the Ricci scalar, and 
$\kappa \equiv 8\pi G$.}
\begin{equation}\label{1}
S_{EH}=\int d^4 x\sqrt{-g} \left( \frac{R}{2\kappa}+{\cal L}_m 
\right) \;,
\end{equation}
is modified to \cite{Capozzielloetal,CDTT}
\begin{equation}\label{2}
S =\int d^4 x\sqrt{-g} \left[ \frac{f(R)}{2\kappa}+{\cal L}_m 
\right] \;,
\end{equation}
where $f(R)$ is an ({\em a priori}) arbitrary function of 
$R$, and the modifications are designed to affect cosmological 
scales and stay tiny at smaller scales in 
order not to violate the  Solar System constraints \cite{Will}. 
The prototype $f(R)=R-\mu^4/R $ (with $\mu \sim 
H_0\sim 
10^{-33}$~eV) is now regarded as an 
unviable toy model at best because it is subject to a violent 
instability \cite{DolgovKawasaki} and violates the experimental 
constraints \cite{ErickcekSmithKamionkowski}.

 In order to be viable, 
modified gravity theories must be free of short 
time scale instabilities and  ghosts 
\cite{DolgovKawasaki,Odintsovconfirm, Nojiri, instabilities}, 
have  a well-posed Cauchy problem \cite{TremblayFaraoni}, and 
have the correct 
cosmological dynamics including an early inflationary era 
followed by a radiation era, a matter era, and a late 
accelerated era (in many  models there are problems with the 
exit from the radiation era \cite{Amendolaetal}).

Modified $f(R)$  gravity comes in three versions: metric 
formalism, in which the action (\ref{2}) is varied with respect 
to the (inverse) metric tensor $g^{ab}$; Palatini $f(R)$ 
gravity, in which variation is with respect to both $g^{ab}$ and 
an independent, non-metric, connection $\Gamma^a_{bc}$ but the 
matter part of the action does not depend on $\Gamma^a_{bc}$ 
\cite{Palatini}; and metric-affine gravity, in 
which also ${\cal L}_m$  depends on the 
non-metric connection \cite{metricaffine}.  Here we focus on the 
metric approach.

Recently, Bertolami, B\"{o}hmer, Harko, and Lobo (hereafter 
BBHL) 
\cite{BBHL}  put a new twist on $f(R)$ gravity  by 
considering the action
\begin{equation}\label{3}
S =\int d^4 x\sqrt{-g} \left\{ \frac{f_1(R)}{2}+\left[1+\lambda  
f_2(R) \right] {\cal L}_m \right\} \;,
\end{equation}
where $f_{1,2}(R)$ are arbitrary functions of the Ricci curvature 
and $\lambda$ is a small parameter (from now on we follow  
\cite{BBHL} and set $\kappa\equiv 8\pi G=1$). 
The novelty consists of the coupling function $f_2(R)$ which 
adds extra freedom and new features. The field equations are 
\begin{eqnarray}\label{4}
&& f_1'(R) R_{ab}-\frac{ f_1(R)}{2}\, g_{ab} = \nabla_a \nabla_b 
f_1'(R)-g_{ab} \Box f_1'(R) \nonumber \\
&& -2\lambda f_2'(R) {\cal 
L}_m R_{ab} 
+2\lambda \left( \nabla_a\nabla_b -g_{ab}\Box \right) \left( 
{\cal L}_m f_2'(R) \right) \nonumber \\
&& +\left[ 1+\lambda f_2(R) 
\right] T_{ab}^{(m)}  \;,
\end{eqnarray}
where a prime denotes differentiation with respect to $R$, 
$\Box\equiv g^{cd} \nabla_c \nabla_d$, and $
T_{ab}^{(m)}=\frac{-2}{\sqrt{-g} }\, \frac{ \delta \left( 
\sqrt{-g}\, {\cal L}_m \right)}{ \delta g^{ab} } $.
Because of the extra explicit coupling to matter $\lambda 
f_2(R)$, $T_{ab}^{(m)}$ is not covariantly conserved and 
 energy is exchanged between ordinary matter ($T_{ab}^{(m)}$) 
and the ``effective matter'' represented by 
terms in $f_2'(R)$ in eq.~(\ref{4}). $T_{ab}^{(m)}$ obeys  
\cite{BBHL} 
\begin{equation}\label{6}
\nabla^b T_{ab}^{(m)}=\frac{ \lambda f_2'(R)}{1+\lambda 
f_2(R)}\left[ g_{ab} {\cal L}_m -T_{ab}^{(m)} \right] \nabla^b R 
\;.
\end{equation}
The BBHL theory contains intriguing phenomenology: all massive 
particles are subject to an extra force, similar to the one  
arising  in scalar-tensor (ST) gravity following a conformal 
transformation to the Einstein frame 
\cite{Dicke,Mbelek}. In 
Einstein frame ST gravity the extra force is due to 
an ``anomalous'' coupling of the matter Lagrangian  to 
the  Brans-Dicke-like scalar $\phi$, 
\begin{equation}\label{7}
S_{EF} =\int d^4 x\sqrt{-g} \left( \frac{R}{2} 
-\frac{1}{2}\nabla^c\phi\nabla_c\phi 
-\mbox{e}^{-\alpha \, \phi} {\cal L}_m \right) \;. 
\end{equation}
There is, however, an important difference 
between Einstein frame ST gravity and BBHL 
theory: while in the former the units of time, length, and mass 
are not constant but scale with appropriate powers of the 
conformal factor of the conformal transformation defining the 
Einstein frame (as explained in \cite{Dicke} and discussed 
extensively in \cite{FaraoniNadeau}), in the latter there 
is no such 
scaling of units.  For this 
reason, the BBHL theory \cite{BBHL} 
can not be reduced to ``ordinary'' ST or string 
gravity (in this respect, string 
theory has the same phenomenology of ST 
gravity---indeed, the low-energy limit of the bosonic string 
is a Brans-Dicke theory with Brans-Dicke parameter $\omega=-1$ 
\cite{bosonicstring}).

The extra force on massive particles generated by the $\lambda 
f_2(R)$ coupling is always present and causes a deviation from 
geodesic paths; therefore, 
massive test particles simply do not exist. Due to this extra 
force, the acceleration 
law in the weak-field limit of BBHL theory assumes  a form  
similar to the one of Modified Newtonian Dynamics (MOND) 
\cite{MOND}, which was originally 
proposed to explain galactic rotation curves 
without  dark matter. MOND has recently 
received a relativistic formulation in the rather complicated 
Tensor-Vector-Scalar (TeVeS) theory of~\cite{TeVeS}. 
The BBHL proposal exhibits MOND-like phenomenology  but has a  
simpler formal structure than TeVeS: as shown below, it 
amounts to a ST theory with two coupling functions, one of which 
is the coupling of the scalar degree of freedom $\phi=R$ to 
matter (unorthodox in ST gravity \cite{mybook}). Of 
course, in order to be viable, the BBHL theory 
must pass the tests mentioned above for $f(R)$ gravity and it 
is not clear yet whether this is possible. In this paper we 
study one of these criteria, namely the stability of the theory 
with respect to local perturbations. In pure $f(R)$ 
gravity, a fatal instability  develops 
on  time scales $ \sim 10^{-26}$~s 
\cite{DolgovKawasaki} when $f''(R)<0$. This 
instability, which we refer to as the ``Dolgov-Kawasaki 
phenomenon'', was discovered in the prototype model 
$f(R)=R-\mu^4/R$ \cite{DolgovKawasaki} which is ruled out 
(and only cured by adding extra terms to $f(R) $
\cite{Odintsovconfirm,Nojiri,Shahn}), and then 
generalized to 
arbitrary $f(R)$ models \cite{mattmodgrav}. For the BBHL 
theory, the corresponding stability criterion turns out to be
$ f_1''(R)+2\lambda f_2''(R) \geq 0$ (see  Sec.~3; see 
Refs.~\cite{otherinstabilities} for other types of 
instabilities).

\section{Equivalence with an anomalous ST theory}

It is well known that pure $f(R)$ gravity (\ref{1})  is 
equivalent to a ST theory 
\cite{STequivalence}; here we revisit this equivalence and  
generalize it to BBHL theory. 

By introducing a new field 
$=\phi$, the action~(\ref{3}) is written as 
\begin{equation} \label{100}
S\int d^4x \sqrt{-g} \left\{ \frac{f_1(\phi)}{2} +\frac{1}{2} \, 
\frac{df_1}{d\phi} \left( R-\phi \right) +\left[ 1+\lambda f_2( 
\phi) \right] {\cal L}_m \right\}
\end{equation}
and, further introducing the field $
\psi(\phi) \equiv f_1'(\phi) $ 
(where now a prime  denotes 
differentiation with respect  to $\phi$ \footnote{This is not an 
abuse of  notations because $\phi=R$.}), one can write
\begin{equation} \label{300}
S=\int d^4x \sqrt{-g} \left[ \frac{\psi R }{2} -V(\psi)\, 
+U(\psi) {\cal L}_m \right] \;,
\end{equation}
where
\begin{eqnarray}
V(\psi) &=& \frac{\phi(\psi) f_1' \left[ \phi (\psi ) \right] 
-f_1\left[ \phi( \psi ) \right] }{2} \;, \label{400}\\
U( \psi) & =& 1+\lambda f_2\left[ \phi( \psi ) \right] 
\;,\label{500}
\end{eqnarray}
with $\phi (\psi)$ given by inverting $ \psi(\phi) \equiv 
f_1'(\phi) $. The 
actions~(\ref{3}) and (\ref{300}) are equivalent when $f_1''(R) 
\neq 0$: in fact, by setting $\phi=R$, eq.~(\ref{300}) reduces 
trivially to eq.~(\ref{3}). Vice-versa, variation of~(\ref{100}) 
with respect to $\phi$ yields  
\begin{equation} \label{600}
\left( R-\phi \right) f_1''(\phi)+2\lambda f_2'(\phi) {\cal 
L}_m=0 \;.
\end{equation}
In vacuo (${\cal L}_m=0$), this equation yields $\phi=R$ 
whenever $f_1''\neq 0$ \cite{STequivalence}. In the presence of 
matter there seem to be other possibilities  which are, 
however, excluded as  follows. When ${\cal L}_m\neq 0$, the 
actions~(\ref{3}) and (\ref{100}) are equivalent if $\left( 
R-\phi \right) f_1''(\phi) +2\lambda f_2''(\phi) {\cal L}_m \neq 
0$. When eq.~(\ref{600}) is satisfied, we have a pathological 
case which, upon integration of this equation, corresponds to 
\begin{equation} \label{700}
\lambda f_2(\phi) {\cal L}_m= \frac{ f_1'(\phi)}{2} \left( 
\phi-R \right) -\frac{ f_1(\phi)}{2} \;. 
\end{equation}
But if eq.~(\ref{700}) holds, then the action~(\ref{100}) 
reduces to 
pure matter without the gravity sector and  we dismiss 
this case. Then, the 
actions~(\ref{3}) and (\ref{300}) are equivalent when $f_1''(R) 
\neq 0$, as in pure $f(R)$ gravity \cite{STequivalence}.  The 
action~(\ref{300}) corresponds to a Brans-Dicke theory 
\cite{BransDicke} with a single scalar field, 
vanishing Brans-Dicke parameter $\omega$, and an unorthodox 
coupling $U(\psi)$ to matter. Actions of 
this kind have been contemplated 
before \cite{Shapiro,Flanagan,SFL}, 
but little is known about them.

\section{BBHL theory and instabilities}

The trace of the field equations is, in terms of $R$, 
\begin{eqnarray}
&& 3\left[ f_1''(R)+2\lambda {\cal L}_m f_2''(R) \right] \Box R 
 +3\left[ f_1'''(R)+2\lambda {\cal L}_m f_2''(R) \right] 
\nonumber \\
&& \nabla^c R \nabla_c R 
+12 \lambda f_2''(R) \nabla^c 
{\cal L}_m  \nabla_c R 
+ f_1'(R) R -2f_1(R)  \nonumber \\
&& +2\lambda {\cal L}_m 
f_2'(R) R  =\left[ 1+\lambda f_2(R) \right] T^{(m)}-6\lambda 
f_2'(R) 
\Box  {\cal L}_m \;, \nonumber \\
&& \label{8}
\end{eqnarray}
where  $T^{(m)}\equiv 
{T^{(m) \,a}}_a$ . As customary in $f(R)$ 
gravity, we parametrize the function $f_1(R)$ as 
$f_1(R)=R+\epsilon \varphi(R)$, where $\epsilon$ and $\lambda$ 
must necessarily be small to respect the 
Solar System constraints \cite{BertottiIessTortora}. 
Following \cite{DolgovKawasaki}, we expand the spacetime 
quantities of interest as the sum of  a background with constant 
curvature and  a small perturbation: $
R=R_0+R_1 $, $T=T_0+T_1$, $ {\cal L}_m={\cal 
L}_0+{\cal L}_1 $,  and the spacetime metric can {\em locally} 
be  approximated 
by $g_{ab}=\eta_{ab}+h_{ab}$, where $\eta_{ab}$ is the Minkowski 
metric. There are really two approximations here. 
The first is an adiabatic expansion around a de Sitter space 
with constant curvature, which is justified on timescales much 
shorter than the Hubble time. The second is a local expansion 
over small spacetime regions that are locally flat (hence the 
appearance of $\eta_{ab}$). These approximations are common in 
$f(R)$ gravity ({\em e.g.}, \cite{DolgovKawasaki,mattmodgrav}) 
and in 1980s literature on inflation.  Accordingly, $
f_1(R) =  R_0+R_1+\epsilon \varphi(R_0)+\epsilon \varphi'(R_0) 
R_1+\, ... $, 
$ f_1'(R) = 1+ \epsilon \varphi '(R_0) + \epsilon 
\varphi''(R_0) R_1+\, ... $ and the linearized version of 
the trace equation~(\ref{8}) in the perturbations becomes
\begin{eqnarray}
&& 3\left[ \epsilon \varphi ''(R_0) +2\lambda f_2''(R_0) \right] 
\Box R_1
+\left[ \epsilon \varphi ''(R_0) R_0 -1 \right. \nonumber \\
&& \left. -\epsilon 
\varphi ' (R_0) 
+2\lambda f_2'(R_0) {\cal L}_0 +2\lambda {\cal L}_0 f_2''(R_0) 
R_0 \right. \nonumber \\
&& \left. -\lambda f_2'(R_0) T_0 
  +6\lambda f_2''(R_0) \left( \Box 
{\cal L}_0 \right) \right] R_1 \nonumber \\
&& = -2 \lambda 
f_2'(R_0) R_0 {\cal L}_1 +\left[ 1+\lambda 
f_2(R_0)  \right] T_1 \nonumber \\
&&  -6\lambda f_2'(R_0) \Box {\cal L}_1 \;, 
\label{9}
\end{eqnarray}
where the zero order equation
\begin{eqnarray}
&& f_1'(R_0) R_0 -2f_1(R_0) +2\lambda {\cal L}_0 f_2'(R_0) 
R_0=\left[ 1+ \lambda f_2(R_0) \right] T_0  \nonumber \\
&& \label{10} 
\end{eqnarray}
has been used. Eq.~(\ref{9}) is further rewritten as 
\begin{eqnarray} 
&& \ddot{R}_1-\nabla^2 R_1 +m_{eff}^2 R_1 
 =
\left\{ 3\left[ \epsilon  \varphi''(R_0) +2\lambda f_2''(R_0) 
\right] \right\}^{-1} \nonumber \\
&& \left\{ 2\lambda  f_2'(R_0){\cal L}_1 -\left[ 1 + \lambda 
f_2(R_0) \right] T_1  +6 \lambda f_2'(R_0) \Box {\cal L}_1  
\right\} 
\end{eqnarray}
where the effective mass $m_{eff} $ of the dynamical degree of 
freedom $R_1$ is given by 
\begin{eqnarray*}
&& m_{eff}^2 = 
\left\{ 3\left[ \epsilon \varphi ''(R_0) +2\lambda f_2''(R_0)\right] 
\right\}^{-1} \left\{
1+ \epsilon \varphi ' (R_0) \right. \nonumber \\
&& \left. + \epsilon \varphi ''(R_0) 
R_0  
-2\lambda {\cal L}_0  \left[ f_2'(R_0) + f_2''(R_0) 
R_0 \right] + \lambda f_2'(R_0) T_0 \right\}
\;.  
\end{eqnarray*} 
The dominant term on the right hand side is 
 $ \left\{ 3\left[ \epsilon \varphi ''(R_0) +2\lambda 
f_2''(R_0)\right] \right\}^{-1} $ and the effective mass squared 
must be 
non-negative for stability. Therefore, 
$ \epsilon \varphi ''(R)+2\lambda f_2''(R) \geq 0 $ 
is the stability criterion for the BBHL theory 
against Dolgov-Kawasaki instabilities.

\section{Outlooks}

The inequality $\epsilon \varphi ''(R)+2\lambda f_2''(R) \geq 0 
$  generalizes the stability condition 
$f''(R) =\epsilon \varphi ''(R) \geq 0$ found in pure $f(R)$ 
gravity \cite{mattmodgrav,SongHuSawicky}.  The survival of BBHL 
theory  \cite{BBHL} is  subject to satisfying the  other 
(physically independent) viability  criteria mentioned above, 
which require a separate analysis and will be analyzed in future 
publications.

\begin{acknowledgments}
We thank Francisco Lobo for a discussion and the Natural 
Sciences and  Engineering Research Council of Canada ({\em 
NSERC}) for financial support.
\end{acknowledgments}


\end{document}